\documentclass{aa}  
\usepackage{natbib}
\usepackage{braket}
\usepackage{graphicx}
\usepackage{booktabs}
\usepackage{setspace}
\usepackage[colorlinks=true,linkcolor=blue,citecolor=blue, urlcolor=blue]{hyperref}
\usepackage{txfonts}
\newcommand\HI{H {\small{I}}}

\newcommand\Lyaf{Ly$\alpha$ forest}
\usepackage{color}
\begin{document}

   \title{Field-level Lyman-alpha forest modeling in redshift space via augmented non-local Fluctuating Gunn-Peterson Approximation}

   \titlerunning{Field-level Lyman-alpha forest via non-local FGPA}

   \author{F. Sinigaglia      \inst{1,2,3,4}\fnmsep\thanks{\email{francesco.sinigaglia@phd.unipd.it}}
          \and
          F.-S. Kitaura\inst{1,2}\fnmsep\thanks{\email{fkitaura@iac.es}}
          \and
          K. Nagamine\inst{5,6,7}
          \and
          Y. Oku\inst{5}
          \and
          A. Balaguera-Antolínez\inst{1,2}
          }

    \authorrunning{F. Sinigaglia et al.}

   \institute{Instituto de Astrof\'isica de Canarias, Calle Via L\'actea s/n, E-38205, La  Laguna, Tenerife, Spain
        \and
        Departamento  de  Astrof\'isica, Universidad de La Laguna,  E-38206, La Laguna, Tenerife, Spain
        \and
        Department of Physics and Astronomy, Università degli Studi di Padova, Vicolo dell’Osservatorio 3, I-35122, Padova, Italy
         \and
        INAF - Osservatorio Astronomico di Padova, Vicolo dell’Osservatorio 5, I-35122, Padova, Italy
        \and
        Theoretical Astrophysics, Department of Earth and Space Science, Graduate School of Science, Osaka University, 1-1 Machikaneyama, Toyonaka, Osaka 560-0043, Japan
        \and
        Kavli-IPMU (WPI), University of Tokyo, 5-1-5 Kashiwanoha, Kashiwa, Chiba, 277-8583, Japan
        \and
        Department of Physics \& Astronomy, University of Nevada, Las Vegas, 4505 S. Maryland Pkwy, Las Vegas, NV 89154-4002, USA
             }

   \date{Received \today; accepted XYZ}

 
  \abstract
   {Devising fast and accurate methods to predict the Lyman-alpha forest at the field level avoiding the computational burden of running large-volume cosmological hydrodynamic simulations is of fundamental importance to quickly generate the massive set of simulations needed by the state-of-the-art galaxy and \Lyaf{} spectroscopic surveys.}
   {We present an improved analytical model to predict the \Lyaf{} at the field level in redshift space from the dark matter field, expanding upon the widely-used Fluctuating Gunn-Peterson approximation (FGPA). Instead of assuming a unique universal relation over the whole considered cosmic volume, we introduce the dependence on the cosmic web environment (knots, filaments, sheets, voids) in the model, thereby effectively accounting for non-local bias. Furthermore, we include a detailed treatment of velocity bias in the redshift space distortions modeling, allowing the velocity bias to be cosmic-web dependent.}
   {We first map the dark matter field from real to redshift space through a particle-based relation including velocity bias, depending on the cosmic web classification of the dark matter field in real space. We then formalize an appropriate functional form for our model, building upon the traditional FGPA and including a cut-off and a boosting factor mimicking a threshold and inverse-threshold bias effect, respectively, with model parameters depending on the cosmic web classification in redshift space. Eventually, we fit the coefficients of the model via an efficient Markov-chain Monte Carlo scheme.}
   {We find evidence for a significant difference of the same model parameters in different environments, suggesting that for the investigated setup the simple standard FGPA is not able to adequately predict the \Lyaf{} in the different cosmic web regimes. We reproduce the summary statistics of the reference cosmological hydrodynamic simulation we use for comparison, yielding accurate mean transmitted flux, probability distribution function, 3D power spectrum, and bispectrum. In particular, we achieve maximum deviation and average deviations accuracy in the \Lyaf{} 3D power spectrum of $\sim 3\%$ and $\sim 0.1\%$ up to $k\sim 0.4 \, h \, {\rm Mpc}^{-1}$, $\sim 5\%$ and $\sim 1.8\%$ up to $k \sim 1.4 \, h \, {\rm Mpc}^{-1}$ .}
   {Our new model outperforms previous analytical efforts to predict the \Lyaf{} at the field level in all the probed summary statistics, and has the potential to become instrumental in the generation of fast accurate mocks for covariance matrices estimation in the context of current and forthcoming \Lyaf{} surveys}


   \maketitle 
%
\section{Introduction}

The neutral hydrogen (\HI{}) absorption features imprinted in quasar spectra, also known as the Lyman-$\alpha$ forest (\Lyaf{}, hereafter), represent a promising cosmological probes over the current and next decades. The \Lyaf{} traces the density field continuously along the line-of-sight, and the high number density of quasar sight lines per square degree delivered by state-of-the-art cosmological redshift surveys will allow building accurate three-dimensional maps of the Universe at $z\gtrsim 1.8$, i.e at distances which are currently beyond the reach of wide-field galaxy redshift surveys. In the await of next-generation astronomical facilities such as the Square Kilometre Array Observatory\footnote{SKAO: \url{https://www.skao.int}} and the Extremely Large Telescope\footnote{ELT: \url{https://elt.eso.org}}, the \Lyaf{} is anticipated to play a pivotal role in refining the constraints on cosmological models and advancing our comprehension of the high-redshift Universe.  

To maximize the scientific return and optimize the extraction of cosmological information from the incoming unprecedented \Lyaf{} data sets provided by surveys as DESI \citep[][]{Levi2013}, Euclid \citep[][]{Amendola2018}, WEAVE-QSO \citep[][]{Pieri2016}, and Subaru-PFS \citep[][]{Takada2014}, it is extremely important to develop accurate analytical models and numerical tools to efficiently interpret and analyze \Lyaf{} observables. 

In particular, a primary goal is to formalize an effective model of the bias relation linking the \Lyaf{} to the dark matter field. In fact, an effective bias mapping allows the prediction of the \Lyaf{} from the dark matter field in a fast and accurate way, enabling the generation of massive sets of \Lyaf{} mock lightcones. Mock catalogs have become the standard tool used in large-scale cosmological surveys to robustly address uncertainties on cosmological parameters, to test pipelines and commissioning tools, and to assess the feasibility of studies targeting new observables, among others. Furthermore, knowing an effective bias description make it possible to forward-model the \Lyaf{} at the field level and iteratively reconstruct the \Lyaf{} \citep[see e.g.][]{Horowitz2019,Porqueres2020} and its Baryon Acoustic Oscillations (BAOs), improving on methods de-evolving cosmic structures back in time with some approximation for the displacement field \citep[e.g.][and later works implementing higher-order Lagrangian Perturbation Theory refinements]{Eisenstein2007}. 

In addition, the dark matter field -- \Lyaf{} connection has found an interesting application in tomographic studies as well, such as in the LATIS survey \citep{Newman2020}, in the CLAMATO survey \citep{Lee2018,Horowitz2022}, and in the PSF Galaxy Evolution Survey \citep{Greene2022}.

Pioneering studies \citep[][]{HuiGnedin1997,Rauch1998,Croft1998,Weinberg1999} proposed to model the \Lyaf{} using a deterministic scaling relation mapping -- known as \textit{Fluctuating Gunn-Peterson approximation} (FGPA, hereafter) -- between the dark matter density and the \HI{} optical depth. While useful, fairly accurate on large linear scales, and fast to compute, the FGPA has been shown to lose accuracy and not adequately model the summary statistics of the \Lyaf{} in the weakly non-linear regime ($k>0.1\, h\,{\rm Mpc}^{-1}$) \citep[e.g.][]{Sinigaglia2022}. A plethora of works employed the FGPA to forward-model the \Lyaf{} from dark matter density fields obtained through N-body simulations \citep[][]{Meiksin2001,Viel2002,Slosar2009,White2010,Rorai2013,Lee2014}, approximated gravity solvers \citep[][]{Horowitz2019} and Gaussian or lognormal random fields \citep[][]{LeGoff2011,FontRibera2012,Farr2020}. In particular, the \textsc{LyAlpha-Colore} method \citep[][]{Farr2020}, applying the FGPA on lognormal fields, with free parameters constrained so as to match the position and width of the acoustic peak in the \Lyaf{} two-point correlation function and the amplitude of the 1D \Lyaf{} power spectrum, was adopted to produce the mock lightcones used for the BAO measurements of the \Lyaf{} from the final eBOSS data release (SDSS DR16) \citep[][]{Bourboux2020}.

On the other hand, more sophisticated iterative methods to model the \Lyaf{} and employing different strategies have been proposed, such as e.g. \textsc{LyMas} \citep[][]{Peirani2014,Peirani2022} and the \textsc{Iteratively-Matched Statistics} \citep[\textsc{Ims},][]{Sorini2016}. These techniques have been shown to yield promising results, although still feature deviations of order $5-20\%$ in the 3D power spectrum. Moreover, those techniques were not applied to approximated gravity solvers, but rely on full N-body simulations. Therefore, they are not able to overcome the computational burden of running massive sets of simulations. 

With the flourishing of machine learning methods, and the extraordinary attention that this field is receiving in astrophysics, the generation of AI\footnote{Artificial Intelligence.}-accelerated simulated cosmological volumes is witnessing a rapid expansion. 

The machine learning method \textsc{bam} \citep[][]{Balaguera2018,Balaguera2019,Pellejero2020,Kitaura2022} and its extension to hydrodynamics \textsc{hydro-bam} \citep[][]{Sinigaglia2021,Sinigaglia2022}, which combines the latest version of a bias mapping method with a physically-motivated strongly-supervised learning strategy and the exploitation of the hierarchy of baryon quantities, has been shown to model two- and three-point statistics to $1\%$ and $\sim 10\%$ level of accuracy, respectively. Therefore, this technique represents a promising way forward to produce \Lyaf{} mock catalogs for the next-generation surveys \citep[see e.g.,][for a concrete application to halo mock catalogs generation for the DESI survey]{Balaguera2023}. 

Deep learning has also achieved competitive results. \citet{Harrington2021} employed deep convolutional generative adversarial networks to learn how to correct the FGPA feeding the density and velocity DM fields as inputs. In a companion paper, \citet{Horowitz2021} exploited the idea of image generation behind deep generative methods and built a conditional convolutional auto-encoder, able to sample representations in latent space of the hydrodynamic fields of a simulation and synthesize the \Lyaf{} with a deep posterior distribution mapping. 

Nonetheless, machine learning techniques to date still require an adequately large training set in order to appropriately address the issue of overfitting, making it still expensive in the case of predicting the \Lyaf{}.   

In this work we propose to elaborate on the FGPA formalism, augmenting it in light of the recent findings on the importance of accounting for long-range and short-range non-local terms in the mapping of dark matter tracers \citep[see e.g.][]{Balaguera2019,Kitaura2022,Sinigaglia2021}. In particular, we explicitly introduce the dependence on the cosmic web environments through the cosmic web classification \citep{Hahn2007} in the FGPA modeling. We showcase the application of our algorithm to map the \Lyaf{} on a few Mpc cells mesh in redshift space, and assess its accuracy by evaluating the deviation of relevant summary statistics (mean transmitted flux, probability distribution function, power spectrum, and bispectrum) of the predictions from the ones of a reference cosmological hydrodynamic simulation. 

The paper is organized as follows. \S\ref{sec:refsim} introduces the cosmological hydrodynamic simulation we use to validate our method. \S\ref{sec:cwc} summarizes the cosmic web classification and its connection to non-local bias. In \S\ref{sec:fgpa} we review the FGPA and describe our improvements in modeling redshift-space distortions (RSD hereafter) and non-local bias terms. \S\ref{sec:results} presents the analysis, results of our predictions and a discussion of them. In \S\ref{sec:lightcone} we describe out to apply our machinery to the generation of \Lyaf{} simulations with lightcone geometry. \S\ref{sec:outlooks} discusses potential future improvements for our model. We conclude in \S\ref{sec:conclusions}.


\section{Reference cosmological hydrodynamic simulation}\label{sec:refsim}

In this section, we present our reference cosmological hydrodynamic simulation.

The reference simulation has been run with the cosmological smoothed-particle hydrodynamics (SPH) code \texttt{GADGET3-OSAKA} \citep{Aoyama2018, Shimizu2019}, a modified version of \texttt{GADGET-3} and a descendant of the popular $N$-body/SPH code \texttt{GADGET-2} \citep{Springel2005}. 
It embeds a comoving volume $V=(500h^{-1}\text{Mpc})^3$ and $N=2\times1024^3$ particles of mass $m_{\rm DM}=8.43\times10^9h^{-1}\text{M}_\odot$ for DM particles and $m_{\rm gas}=1.57\times 10^9 h^{-1}\text{M}_\odot$ for gas particles.
The gravitational softening length is set to $\epsilon_g = 16 h^{-1}$\,kpc (comoving), but we allow the baryonic smoothing length to become as small as $0.1\epsilon_g$. This means that the minimum baryonic smoothing at $z=2$ is about physical $533\,h^{-1}$\,pc. 
The star formation and supernova feedback are treated as described in \citet{Shimizu2019}.
The code contains also important refinements, such as the density-independent formulation of SPH and the time-step limiter \citep[][]{Saitoh2009, Saitoh2013, Hopkins2013}. 

The main baryonic processes which shape the evolution of the gas are photo-heating, photo-ionization under the UV background radiation \cite[][]{Haardt2012}, and radiative cooling. All these processes are accounted for and solved by the \texttt{Grackle} library\footnote{\url{https://grackle.readthedocs.io/}} \citep{Smith2017}, which determines the chemistry for atomic (H, D and He) and molecular (H$_2$ and HD) species. The chemical enrichment from supernova is also treated with the \texttt{CELib} chemical evolution library by \citet{Saitoh2017}.
The initial conditions are generated at redshift $z=99$ using \texttt{MUSIC2} \citep{Hahn2021} with cosmological parameters taken from \cite{Planck2018}. 



In this work we use the output at $z=2$ (for which the computation times amount to $\sim 1.44\times10^5$ CPU hours), reading both gas and dark matter properties. 

To compute the \Lyaf{} flux field $F=\exp(-\tau)$\footnote{The notation $F=\exp(-\tau)$ for the \Lyaf{} flux field actually refers to the transmitted flux $F/F_{\rm c}$, i.e. to the quasar spectrum normalized to its continuum $F_{\rm c}$. We will omit hereafter the normalization and denote the transmitted flux just as $F$ unless specified otherwise.}, we first obtain the \HI{} optical depth $\tau$ by means of a line-of-sight integration as follows \citep{Nagamine21}: 
\begin{equation}\label{eq:tau}
    \tau = \frac{\pi e^2}{m_e c} \sum_j f \, \phi(x-x_j)\, n_{\rm HI}(x_j)\, \Delta l,
\end{equation}
where $e$, $m_e$, $c$, $n_{\rm HI}$, $f$, $x_j$, $\Delta l$ denote respectively the electron charge, electron mass, speed of light in vacuum, \HI{} number density, the absorption oscillator strength, the line-of-sight coordinate of the $j$-th cell and the physical cell size. The Voigt-line profile $\phi(x)$ in Eq.~(\ref{eq:tau}) is provided by the fitting formula of \cite{Tasitsiomi2006}. Where necessary, relevant quantities (e.g. \HI{} number density) are previously interpolated on the mesh according to the SPH kernel of the simulation. Coordinates $x_j$ of cells along the line-of-sight refer to the outcome of interpolation of particles based either on their positions in real space $r_j$  or redshift-space $s_j=r_j + v^{\rm los}_j/aH$, where $v^{\rm los}_j$, $a$ and $H$ are the $j$-th particle velocity component along the line-of-sight, the scale factor and the Hubble parameter at $z=2$, respectively. 

We interpolate the dark matter and the \Lyaf{} fields in real and in redshift space onto a $256^3$ cells cubic mesh using a CIC mass assignment scheme \citep[][]{1981csup.book.....H}. Such resolution corresponds to a physical cell-volume of $\partial V \sim (1.95 \, h^{-1} \, \rm{Mpc})^{3}$ and a Nyquist frequency of $k_{\rm nyq}\sim1.6\,h\,$Mpc$^{-1}$. 


It is worth noting that the simulations utilized for the analyses of the \Lyaf{} in \citet{Momose2021,Nagamine21}, the exploration of the cosmological \HI{} distribution in \citet{Ando19}, and the examination of bias in cosmological gas tracers in \citet{Sinigaglia2021,Sinigaglia2022} were all executed using the same code.

\section{Cosmic web classification and non-local bias} \label{sec:cwc}


The cosmic web \citep[see e.g.,][]{Bond1996} arises as a result of gravitational instability and the formation and growth of cosmic structures from tiny perturbations of the primordial matter density field in the Early Universe. While several different methods have been proposed in the literature to mathematically define the cosmic web \citep[see][for a summary]{Cautun2014,Libeskind2017}, we focus here on the following procedure. 

To quantitatively describe the large-scale matter distribution and split it into different cosmic web environments, \cite{Hahn2007} proposed a classification scheme based on the signature of the eigenvalues of the gravitational tidal field tensor
\begin{equation}
\mathcal{T}_{ij}(\vec{r})=\partial_i\partial_j\phi(\vec{r}) \, ,
\end{equation}
where $\phi$ denotes the gravitational potential and $\vec{r}$ stands for Eulerian coordinates. Considering the equations of motion in comoving coordinates 
\begin{equation}
\ddot{\vec{r}} = -\nabla\phi(\vec{r})
\end{equation} 
for a test particle subject to the gravitational potential $\phi$, and assuming $\nabla\phi(\vec{r})=0$ at the center of mass of dark matter haloes (i.e. there is a local minimum), one can linearize the equation of motions and realize that the dynamics close to local extrema of the gravitational potential in the linear regime is ruled by the three eigenvalues $\lambda_i$ ($i=1,2,3$) of $\mathcal{T}_{ij}$ \citep[we refer the reader to][for more details on the calculations]{Hahn2007}. 
In close analogy to Zel'dovich approximation \citep{Zeldovich1970}, and sorting the $\lambda_i$ in decreasing order such that $\lambda_1\ge\lambda_2\ge\lambda_3$, \cite{Hahn2007} defined a region at coordinates $\vec{r}$ to belong to:
\begin{itemize}
    \item a knot, if $\lambda_1,\lambda_2\, \lambda_3\ge 0$;
    \item a filament, if $\lambda_1, \lambda_2\ge 0, \lambda_3< 0$;
    \item a sheet, if $\lambda_1\ge 0, \lambda_2, \lambda_3<0$;
    \item a void, if $\lambda_1, \lambda_2, \lambda_3< 0$.
\end{itemize}

At this point, one can drop the assumption of local extrema at the center of mass of the dark matter halo and generalize the cosmic web classification to any point of the density field. This implies the emergence of a constant additive acceleration term to the linearized equations of motion, which however does not affect the role played by the tidal tensor in linear order. 

Elaborating on this work, \cite{ForeroRomero2009} proposed to relax the threshold to classify the cosmic web, based on simple dynamical arguments and realizing that highering the threshold from $\lambda_{\rm th}=0$ to $\lambda_{\rm th}=0.1$ provides a classification that better matches the visual appearance of the cosmic web. 

The T-web classification has also been shown to provide a direct connection between the phenomenology of the cosmic web and the mathematical description of halo bias relying on Eulerian perturbation theory \citep[see e.g.][]{McDonaldRoy2009}. In fact, discriminating between different T-web environments corresponds to considering a perturbative bias expansion up to third order including both local and non-local long-range bias terms. In this sense, following \cite{Kitaura2022}, let us denote the three invariants of the tidal field tensor as:
\begin{itemize}
    \item $I_1=\lambda_1 + \lambda_2 + \lambda_3$;
    \item $I_2 = \lambda_1\lambda_2 + \lambda_1\lambda_3 + \lambda_2\lambda_3$;
    \item $I_3 = \lambda_1\lambda_2\lambda_3$.
\end{itemize}

These three items represent an alternative formulation of the perturbative bias expansion up to third order \citep{Kitaura2022}, but they can also be straightforwardly linked to the phenomenological T-web classification as follows:
\begin{itemize}
    \item knots: $I_3>0$ \& $I_2>0$ \& $I_1>\lambda_1$;
    \item filaments: $I_3<0 ~ \& ~I_2<0 ~ || ~ I_3<0 ~ \& ~ I_2>0 ~ \& ~ \lambda_3<I_1<\lambda_3-\lambda_2\lambda_3/\lambda_1$;
    \item sheets: $I_3>0 ~ \& ~ I_2<0 ~ || ~ I_3<0 ~ \& ~ I_2>0 ~ \& ~ \lambda_1-\lambda_2\lambda_3/\lambda_1<I_1<\lambda_1$;
    \item voids: $I_3<0$ \, \& \, $I_2>0$ \, \& \, $I_1<\lambda_1$.
\end{itemize}

Eventually, one can generalize this description by relaxing the value of the eigenvalues threshold and replacing zero with any other arbitrary value. 

From an intuitive point of view, this means that T-web models the full perturbative expansion up to the third order adopting a low-resolution binning, i.e. just four categories identified as cosmic web types \citep[see][and references therein]{Kitaura2022}. While using the invariants $I_i$ of $\mathcal{T}_{ij}$ has been shown to provide a more accurate description of non-local bias and of anisotropic clustering than T-web, the T-web classification has the advantage of being much less likely to incur into overfitting.

In this work, we apply the T-web classification to the modeling of the \Lyaf{}, in order to introduce non-local contributions within the FGPA prescription. In particular, as will be described in more detail in \S\ref{sec:fgpa}, we make both the model for RSDs and for the FGPA dependent on the T-web by fitting the parameters of such models separately for each cosmic web type. Also, we extract the cosmic web classification both in real and in redshift space.  Table \ref{tab:tweb} reports the volume filling factor of each cosmic web type for the real space dark matter field, and the redshift space dark matter field obtained by applying the RSD description including velocity bias dependent on the real-space T-web with parameters described in the upper part of Table \ref{tab:table_pars} (see \S\ref{sec:rsd} for the details about the procedure), and used to compute our final non-local FGPA model. One realizes that there is only a tiny  sub-percent difference in volume filling factors between real and redshift space T-web classification, with sheets being the most frequent cosmic web type ($\sim 50\%$ of the volume), knots being the rarest ($\sim 2-3\%$ of the volume), and filaments and voids representing intermediate cases.     

\begin{table}
    \centering
    \begin{tabular}{lcc}
    \toprule
         & Real space $(\%)$ & Redshift space $(\%)$ \\
    \midrule
     Knots    & $2.1$ & $2.7$\\
     Filaments & $21.4$ & $21.9$ \\
     Sheets & $49.2$ & $48.8$ \\
     Voids & $27.3$ & $26.6$\\
    \bottomrule
    \end{tabular}
    \vspace{0.2cm}
    \caption{Volume filling factors of each cosmic web type for the real space dark matter field, and the redshift space dark matter field obtained by applying the RSD description including velocity bias dependent on the real-space T-web with parameters described in the upper part of Table \ref{tab:table_pars} (see \S\ref{sec:rsd} for the details about the procedure), and used to compute our final non-local FGPA model.}
    \label{tab:tweb}
\end{table}

\section{FGPA modeling} \label{sec:fgpa}
 
 
In this section, we first introduce the FGPA approximation in the standard scenario, and present later on our improvement on the modeling within such a framework.

\subsection{Standard FGPA}

A popular way to compute a fast proxy for the \Lyaf{} consists in the FGPA, which assumes equilibrium between optically-thin photoionization and collisional recombination of the intergalactic \HI{}.  
In particular, because the majority \citep[$>90\%$ in volume and $>50\%$ in mass,][]{Lukic2015} of the gas probed by the \Lyaf{} found in regions with mildly non-linear density contrasts is diffuse and is not shock-heated, the gas density $\rho_{\rm gas}$ and temperature $T_{\rm gas}$ can be linked through the power-law relation
\begin{equation}
    T_{\rm gas}=T_0(\rho_{\rm gas}/\bar{\rho}_{\rm gas})^\gamma 
\end{equation}
where $\bar{\rho}$ is mean gas density and $T_0$ and $\gamma$ depend on the reionization history and on the spectral slope of the UV background model. These vary commonly within the ranges $4000\,\rm{K}\lesssim T_{0} \lesssim 10^{4}\,\rm{K}$ and $0.3\lesssim\gamma\lesssim 0.6$ \citep[see e.g.,][]{HuiGnedin1997}. Assuming photo-ionization equilibrium, one can express the \Lyaf{} optical depth $\tau$ as a function of the gas density as 
\begin{equation}\label{eq:fgpa}
\tau = A(1+\delta_{\rm gas})^\alpha .
\end{equation}
This expression represents the FGPA, as $\tau$ is described as a field that fluctuates with the underlying gas distribution. While $\alpha=2-0.7 \, \gamma\sim1.6$ \citep[see e.g.,][]{Weinberg1999, Viel2002,Seljak2012,Rorai2013,Lukic2015,Cieplak2016,Horowitz2019}, $A$ is a normalization constant which depends on redshift and on the details of the hydrodynamics \citep[e.g.][]{Weinberg1999}. Given that in the cool low-density regions, dark matter and gas densities display a very high cross-correlation \citep[see e.g.][]{Sinigaglia2021} and that the FGPA consists in a useful tool to be applied to the dark matter field to obtain \Lyaf{} predictions without solving the equations of hydrodynamics, $\delta_{\rm gas}$ can be replaced with $\delta_{\rm dm}$ in Eq.~(\ref{eq:fgpa}), so that the FGPA is applied directly to the dark matter field:
\begin{equation}\label{eq:fgpa_dm}
\tau = A(1+\delta_{\rm dm})^\alpha .
\end{equation}

The FGPA has been shown to represent a good approximation for very high-resolution full N-body simulations \citep[e.g.][]{Sorini2016,Kooistra2022}, i.e. when representing density fields on regular grids with cell size $l\sim 150-200 \, {\rm kpc} \, h^{-1}$, where particle positions and velocities are known with the high accuracy thanks to the exact solution of collisionless and fluid dynamics for dark matter and gas particles/cells, respectively.  However, this is not the case if either a coarser grid resolution ($\gtrsim 1-2 \, \rm{Mpc}\, h^{-1}$) and/or approximated gravity solvers are to be adopted, as in the case of the massive generation of large-volume \Lyaf{} boxes encompassing all the universe up to $z\sim 4$ \citep[e.g.][]{Farr2020}. In particular, the summary statistics predicted by the FGPA have been shown to depart from the corresponding hydrodynamic flux field statistics obtained with the full hydro computation already at cell resolutions of $l\sim 0.8 \, \rm{Mpc}\, h^{-1}$, both in real and in redshift space \citep{Sinigaglia2022}.

In this work, we compute the FGPA based on the cubic mesh used to CIC-interpolate the simulation particles, i.e. with physical cell size $l\sim 1.95 \, {\rm Mpc} \, h^{-1}$. To motivate the choice of a coarse grid, we notice that such resolution implies that one needs to resort to a grid with $N=5120^3$ cells to represent a simulation box of volume $V=(10\,{\rm Gpc\, h^{-1}})^3$, covering a full-sky cosmic realization at $0\le z\lesssim 3.8$, to cover the relevant volume needed for \Lyaf{} studies. This implies already a quite heavy computational burden, especially when hundreds of mock lightcones are to be generated, and resorting to higher resolution meshes would make the process practically unfeasible.  

To put ourselves in realistic observational conditions, we compute the FGPA in redshift space. To this end, we displace particles from real to redshift space via the mapping \citep{Kaiser1987,Hamilton1998}
\begin{equation}\label{eq:rsd_basic}
    \vec{s}_i = \vec{r}_i + \frac{\left(\vec{v_i}\cdot \hat{\vec{r}_i}\right)\hat{\vec{r}}_i}{aH} \,
\end{equation}
where $\vec{s}_i$ and $\vec{r}_i$ are the Eulerian comoving coordinates of the $i$th particle in redshift space and real space, $\hat{\vec{r}}_i=\vec{r}_i/|\vec{r}_i|$, $\vec{v}_i$ is its velocity, and $a$ and $H$ are the scale factor and the Hubble parameter at the redshift of interest. respectively. Under the assumption of plane-parallel approximation that we adopt, and therefore neglecting the lightcone geometry, Eq.~(\ref{eq:rsd_basic}) simplifies to only one scalar component along the line of sight. 

In this setup, we compute the \textit{standard}\footnote{We adopt the nomenclature \textit{standard} in contrast to the non-local FGPA version presented in the following sections.} FGPA by setting $\alpha=1.6$ and determining the parameter $A$ as the value which matches the normalization of the \Lyaf{} 3D power spectrum on large scales. In this case, we find $A=0.27$. 

\subsection{RSD modeling}\label{sec:rsd}

While the mapping from real to redshift space presented in Eq.~(\ref{eq:rsd_basic}) holds exactly for dark matter particles, it does not account for any velocity bias contribution in the \Lyaf{}. Therefore, following \cite{Sinigaglia2022}, we generalize such a model as follows.   

We start by considering the dark matter field in real space. For each cell $i$, we assign a set of $N$ fictitious particles, with Eulerian position $\vec{r}_i$ coinciding with the center of the cell. Applying an inverse nearest grid point (NGP) scheme, each particle $j$ inside cell $i$ is assigned a mass $M_i=\rho_i V_i/N$, where $\rho_i$ and $V_i$ are the dark matter density and volume of the cell, respectively. The $j$th particles is then displaced from real to redshift space using a modified version of Eq.~(\ref{eq:rsd_basic}) accounting for velocity bias:
\begin{equation}\label{eq:rsd}
    \vec{s}_j = \vec{r}_j + b_v\,\frac{(\vec{v}_{{\rm dm},j} \cdot \hat{\vec{r}}_j)\,\hat{\vec{r}}_j}{aH} \, ,
\end{equation}
where $\vec{s}_j$ and $\vec{r}_j$ are redshift space and real space Eulerian comoving coordinates of the $j$th particle, $\vec{v}_{{\rm dm},j}$ is the modeled velocity of the particle, and $b_v$ is a velocity bias factor. The particle velocity $\vec{v}_{{\rm dm},j}$ is modeled as the sum of two components $\vec{v}_{{\rm dm},j}=\vec{v}^{\rm coh}_{{\rm dm},j}+\vec{v}^{\rm disp}_{{\rm dm},j}$. Here, $\vec{v}^{\rm coh}_{{\rm dm},j}=\vec{v}^{\rm sim}_{{\rm dm},i}$ corresponds to the velocity field at cell $i$, interpolated on the mesh with the same mass assignment scheme as the density field, and models the large-scale coherent flows. We draw the attention of the reader to the fact that the velocity bias $b_v$ directly multiplies the coherent flows component of the velocity field. On the other hand, $\vec{v}^{\rm disp}_{{\rm dm},j}$ is built as a velocity dispersion term, randomly sampled from a Gaussian with zero mean and variance proportional to the value of the density field inside the cell $i$ in real space through a power law: $\vec{v}^{\rm coh}_{{\rm dm},j}\curvearrowleft\mathcal{N} \left(\mu=0,\sigma=B\,(1+\delta)^\beta\right)$, with $B$ and $\beta$ free parameters \citep{Hess2013,Kitaura2014}. This latter component models the quasi-virialized velocity dispersion motions, which are smoothed out by the interpolation of the velocity field on the mesh.

To determine the optimal values for the RSDs and FGPA normalization parameters, we jointly maximize the cross-correlation coefficient between the FGPA prediction and the \Lyaf{} flux field from the reference simulation and match the normalization of the \Lyaf{} 3D power spectrum at large scales. After fixing $\alpha=1.6$ as fiducial value, we find the choice $A=0.7$, $b_v=-0.8$, $b_v=1.5$, $\beta=2.0$ to fulfill the aforementioned constraints. 

\subsection{Non-local bias contribution}\label{sec:non-local}

Looking at Eq.~(\ref{eq:fgpa_dm}), one easily realizes that the standard FGPA prescription provides a way to compute the \Lyaf{} optical depth relying exclusively on the local dark matter density, and neglecting any non-local contribution. While the simplicity of the FGPA model is appealing as it allows to compute the \Lyaf{} with just a few parameters, it attempts at modeling distinct cosmic web environments with the same physical model, therefore failing at capturing the intrinsic diversity of physical conditions. As a major contribution of this paper, we propose to make the FGPA cosmic web dependent, to relax the exponent $\alpha$ and let it variable, and to determine the sets of coefficients that best describe each cosmic web type. We point out that we are considering at this stage the dark matter field in redshift space as modeled in Eq.~(\ref{eq:rsd}), and we therefore extract the T-web classification directly in redshift space. 

Analogously, we also make the RSD model in Eq.~(\ref{eq:rsd}) dependent on the cosmic web environment, thereby passing from the $3$ free parameters $\{b_v, B, \beta\}$, to $12$ parameters ($3$ parameters $\times$ $4$ cosmic web environments). However, in order to keep the number of free parameters as small as possible, we notice that the virial theorem allows us to write $v\propto \sqrt{\rho}$, therefore we can fix $\beta=0.5$. Moreover, peculiar velocity are closer to the linear or quasi-linear regime in sheets and voids, hence one can expect the velocity dispersion component to be negligible there, allowing us to set $B_{\rm sh}=B_{\rm vd}=0$ \footnote{We hereafter adopt the following subscript notation: kn=knots, fl=filaments, sh=sheets, vd=voids.}. These two tricks allow us to reduce the number of free parameters from $12$ to $6$. We notice that, in contrast to the cosmic web classification discussed above, here the T-web is extracted in real space, as the T-web itself is needed to perform the mapping from real to redshift space.

In summary, to introduce the non-local dependence on the cosmic web in the FGPA we:
\begin{enumerate}
\item extract the T-web from the dark matter in real space;
\item determine the distinct parameters $\{b_v, B, \beta\}$ depending on the real-space T-web;
\item map particles from real to redshift space adopting the cosmic web dependent version of Eq.~(\ref{eq:rsd});
\item extract the T-web from the dark matter in redshift space;
\item use the redshift-space T-web to model non-localities in the FGPA by making the free parameters $\{A,\alpha\}$ dependent on the cosmic web category each cell belongs to. 
\end{enumerate}


\subsection{Threshold bias}\label{sec:threshold}

It is well known from cosmological hydrodynamic simulations that the \Lyaf{} transmitted flux one-point probability distribution function (PDF) is starkly bimodal, featuring two sharp peaks around $F=0$ and $F=1$, and displaying a much lower occurrence of cells with intermediate values $0.1\lesssim F \lesssim 0.9$ \citep[e.g.,][and references therein]{Nagamine21}. These characteristics can be mainly attributed to the exponential mapping $F=\exp(-\tau)$, which regularizes the domain of $\tau$ mapping it in the interval $\left[0,1\right]$, truncating low values of $\tau$ to $F\sim 1$, as well as rapidly saturating high values of $\tau$ to $F\sim 0$. The bimodal nature of the \Lyaf{} PDF, together with the high non-linearity induced by the exponential mapping, makes it particularly hard to capture such behavior and accurately predict fluxes, which can suffer rapid variations within the domain, even in neighboring cells. 

To improve on the FGPA model regarding this aspect, we adopt the perspective of introducing a threshold bias \citep{Kaiser1984,Bardeen1986,Sheth2001,MoWhite2002,Kitaura2014,Neyrinck2014,Vakili2017}, which in state-of-the-art models of halo bias is used to suppress the formation in overdense regions. 

In analogy with the formalism of the \textsc{Patchy} method \cite{Kitaura2014,Kitaura2015,Kitaura2016b,Kitaura2016,Vakili2017} and expanding upon it, we update Eq.~(\ref{eq:fgpa_dm}) by introducing two multiplicative exponential terms
\begin{equation} \label{eq:fgpa_cutoff}
    \tau = A(1+\delta)^\alpha \exp\left(-\frac{\delta}{\delta^*_1}\right)\exp\left(\frac{\delta}{\delta^*_2}\right) \, ,
\end{equation}
where the term $\exp\left(-\delta/\delta^*_1\right)$ represents a threshold bias term acting as cut-off, while $\exp\left(\delta/\delta^*_2\right)$ an inverse threshold bias acting as a boost. Either term have a negligible effect when $\delta \ll \delta^*$, whereas they acquire importance when $\delta \gtrsim \delta^*$. Here we do not use the step function used in \cite{Vakili2017}, and $\delta^*_1$ and $\delta^*_2$ are left free. Also, the parameters for RSD are left unchanged ($b_v=-0.8$, $b_v=1.5$, $\beta=2.0$). We fix again $\alpha=1.6$, and find as best-fitting parameter values $A=0.26$, $\delta^*_1=0.81$, and $\delta^*_2=0.49$.

On top of this modification, following \S\ref{sec:non-local}, we make the $4$ free parameters of the model depend on the T-web classification, i.e. $\{A,\alpha\,\delta^*_1,\delta^*_2\}_i$, with $i=\{{\rm knots,\, filaments,\, sheets,\, voids}\}$. In this way, we model differently the FGPA depending on distinct cosmic web environments, thereby effectively including non-local bias. 

So far, the model has $22=6+16$ free parameters ($6$ for RSD as described in \S\ref{sec:rsd}, $16=4\times4$ for the augmented non-local FGPA prescription). We describe the procedure we adopt to determine such parameters and our findings in \S\ref{sec:results} and Table \ref{tab:table_pars}, respectively.

\begin{table}
    \centering
    \begin{tabular}{ccl}
    \toprule
    {\bf Parameter} & {\bf Value} & {\bf Description}\\
    \toprule
     & RSD & \\
     \midrule
    $b_{v,{\rm kn}}$     &  {\tiny $-0.9$} & Coh. flows velocity bias in knots \\
     $b_{v,{\rm fl}}$     &  {\tiny $-0.8$} & Coh. flows velocity bias in filaments \\
     $b_{v,{\rm sh}}$     &  {\tiny$-1.$} & Coh. flows velocity bias in sheets \\
     $b_{v,{\rm vd}}$     &  {\tiny$-0.5$} & Coh. flows velocity bias in voids \\
     $B_{\rm kn}$     &  {\tiny$3.0$} & Q.v.m. velocity bias in knots\\
     $B_{\rm fl}$     &  {\tiny$2.0$} & Q.v.m. velocity bias in filaments\\
     $B_{\rm sh}$     &  {\tiny$0.0$} & Q.v.m. velocity bias in sheets\\
     $B_{\rm vd}$     &  {\tiny$0.0$} & Q.v.m. velocity bias in voids\\
     $\beta_{\rm kn}$     &  {\tiny $0.5$} & Q.v.m. exponent in knots\\
     $\beta_{\rm fl}$     &  {\tiny$0.5$} & Q.v.m. exponent in filaments\\
     $\beta_{\rm sh}$     &  {\tiny ---} & Q.v.m. exponent in sheets\\
     $\beta_{\rm vd}$     &  {\tiny ---} & Q.v.m. exponent in voids\\
    \midrule
    & Mean bias & \\
     \midrule
    $A_{\rm kn}$     & {\tiny $0.23^{+0.34}_{-0.19}$} & Normalization in knots \\
    $A_{\rm fl}$     & {\tiny $0.11^{+0.51}_{-0.09}$} & Normalization in filaments \\
    $A_{\rm sh}$     & {\tiny $0.20^{+0.41}_{-0.16}$} & Normalization in sheets \\
    $A_{\rm vd}$     & {\tiny $0.15^{+0.36}_{-0.12}$} & Normalization in voids \\
    $\alpha_{\rm kn}$     & {\tiny $4.03^{+0.66}_{-0.67}$} & Power-law exponent in knots \\
    $\alpha_{\rm fl}$     & {\tiny $3.97^{+0.70}_{-0.69}$} & Power-law exponent in filaments \\
    $\alpha_{\rm sh}$     & {\tiny $2.21^{+0.67}_{-0.71}$} & Power-law exponent in sheets \\
    $\alpha_{\rm vd}$     & {\tiny $2.03^{+0.72}_{-0.68}$} & Power-law exponent in voids \\
    $\delta^*_{1,\rm kn}$     & {\tiny $1.36^{+0.55}_{-0.48}$} & Exponential cut-off scale in knots \\
    $\delta^*_{1,\rm fl}$     & {\tiny $0.93^{+0.51}_{-0.50}$} & Exponential cut-off scale in filaments \\
    $\delta^*_{1,\rm sh}$     & {\tiny $0.85^{+0.42}_{-0.45}$} & Exponential cut-off scale in sheets \\
    $\delta^*_{1,\rm vd}$     & {\tiny $1.01^{+0.39}_{-0.68}$} & Exponential cut-off scale in voids \\
    $\delta^*_{2,\rm kn}$     & {\tiny $0.41^{+0.48}_{-0.37}$} & Exponential boost scale in knots \\
    $\delta^*_{2,\rm fl}$     & {\tiny $0.17^{+0.51}_{-0.11}$} & Exponential boost scale in filaments \\
    $\delta^*_{2,\rm sh}$    & {\tiny $0.60^{+0.42}_{-0.39}$} & Exponential boost scale in sheets  \\
    $\delta^*_{2,\rm vd}$     & {\tiny $0.25^{+0.47}_{-0.18}$} & Exponential boost scale in voids \\
    \midrule
    & Stochasticity & \\
     \midrule 
      $f_\epsilon$ & {\tiny $1.38^{+0.38}_{-0.27}$} & Normalization of stochastic term \\
     $n$ & {\tiny $0.11^{+0.38}_{-0.08}$} & Number of successes\\
     $p$ & {\tiny $0.58^{+0.34}_{-0.28}$} & Success probability ($0<p<1$) \\
    \bottomrule
    \end{tabular}
    \vspace{0.2cm}
    \caption{Best-fit model parameters for our preferred non-local FGPA prescription with stochasticity. The left column reports the symbol used throughout the paper for each parameter, the central column reports the value and the associated uncertainties, and the right column provides a brief description of what the parameter stands for.} 
    \label{tab:table_pars}
\end{table}

\subsection{Stochasticity}\label{sec:stochasticity}

As final ingredient, we add a stochastic component $\epsilon$ to the model expressed in Eq.~(\ref{eq:fgpa_cutoff}) and the subsequent non-local modification:
\begin{equation} \label{eq:fgpa_final}
    \tau = A_i(1+\delta)^{\alpha_i} \exp\left(-\frac{\delta}{\delta^*_{1,i}}\right)\exp\left(\frac{\delta}{\delta^*_{2,i}}\right) + \epsilon_i , 
\end{equation}
where $\epsilon^j_i=f_{\epsilon,i}\, N^j$ is the noise term in the $j$th cell for $i=\{{\rm knots,\, filaments,\, sheets,\, voids}\}$, and $f_{\epsilon,i}$ is a normalization constant. While $f_\epsilon$ is a parameter controlling the amplitude of the stochastic term, $N$ is sampled from a negative binomial distribution \citep[see][]{Kitaura2014,Vakili2017}, parametriszd as\footnote{While other parametrizations are possible, we stick to the one used in this work and reported in the \texttt{numpy} \citep{Numpy} documentation, for the sake of the reproducibility of results.} 
\begin{equation}
    P(N \, | \,  n,p) = \frac{\Gamma(N+n)}{N! \, \Gamma(n)} \, p^n(1-p)^N,
\end{equation}

where $N$ is the sampled term and corresponds to the number of failures in the standard negative binomial distribution definition, $n$ stands for the number of successes\footnote{$n$ is in principle an integer number, but its definition can be generalized to reals, as done in this work.}, $N+n$ is the total number of trials, and $p$ is the success probability. 

In particular, as will be commented more in detail in \S\ref{sec:results}, the latest non-local modification of Eq.~(\ref{eq:fgpa_cutoff}) without stochasticity yields predictions that lack power towards small scales, and display suppression of small-scale structures, especially in the underdense environments. To this end, we decide to model the stochastic term only in voids, leaving the improved non-local FGPA model deterministic in knots, sheets, and voids, to limit the loss of cross-correlation. This implies $f_\epsilon=0$ in knots, filaments, and sheets. As will be shown later in the paper, we find this choice to be sufficient to accomplish the desired goal.

We can develop an intuition about the application of discrete tracers statistics to the \Lyaf{} in voids as follows. From a mathematical point of view, we start by noticing that we can in principle sample the noise term from any arbitrary continuous distribution. In fact, we are effectively restricting our study to the second moment of the distribution. Starting from this, we can then wonder how to model the variance of the distribution. One possibility would be to assume a Poisson variance, or going beyond it, as done in this paper by choosing the negative binomial. Therefore, our model is statistically robust.   
This stochastic component can then be interpreted as follows. Assuming that \HI{} is found in sparse clouds in voids at high redshift, then we can treat them as point sources and model them in a similar way as haloes and galaxies. We will test the validity of this assumption in future work. For the moment, we argue that this effective model works well and achieves the goal it was designed for (see \S\ref{sec:results} for the results).

The negative binomial distribution can be interpreted as an overdispersed alternative to the Poisson distribution, where one can allow the mean and variance to be different. This aspect can turn out to be useful in cases when two events have a positively correlated occurrence (i.e., the two events have a positive covariance term and are not independent), and this causes a larger variance than in the case in which the events are independent. In the presented parametrization, $n$ controls the deviation from Poissonity, making the distribution converge to the Poisson distribution for large $n$ and causing an overdispersion for small $n$. While a thorough investigation of the deviation from Poissonity for the \Lyaf{} goes beyond the scope of this paper, it is sensible and convenient to allow the distribution from which we sample the stochastic term to feature a non-Poissonian variance including a correlation term \citep[see e.g.][for a detailed treatment applied to galaxies]{Peebles1980}.



\section{Analysis, results and discussion}\label{sec:results}


In this section, we describe the procedure we adopt to determine the optimal parameters for our model, and present our findings.

\begin{figure*}
    \centering
    \includegraphics[width=\textwidth]{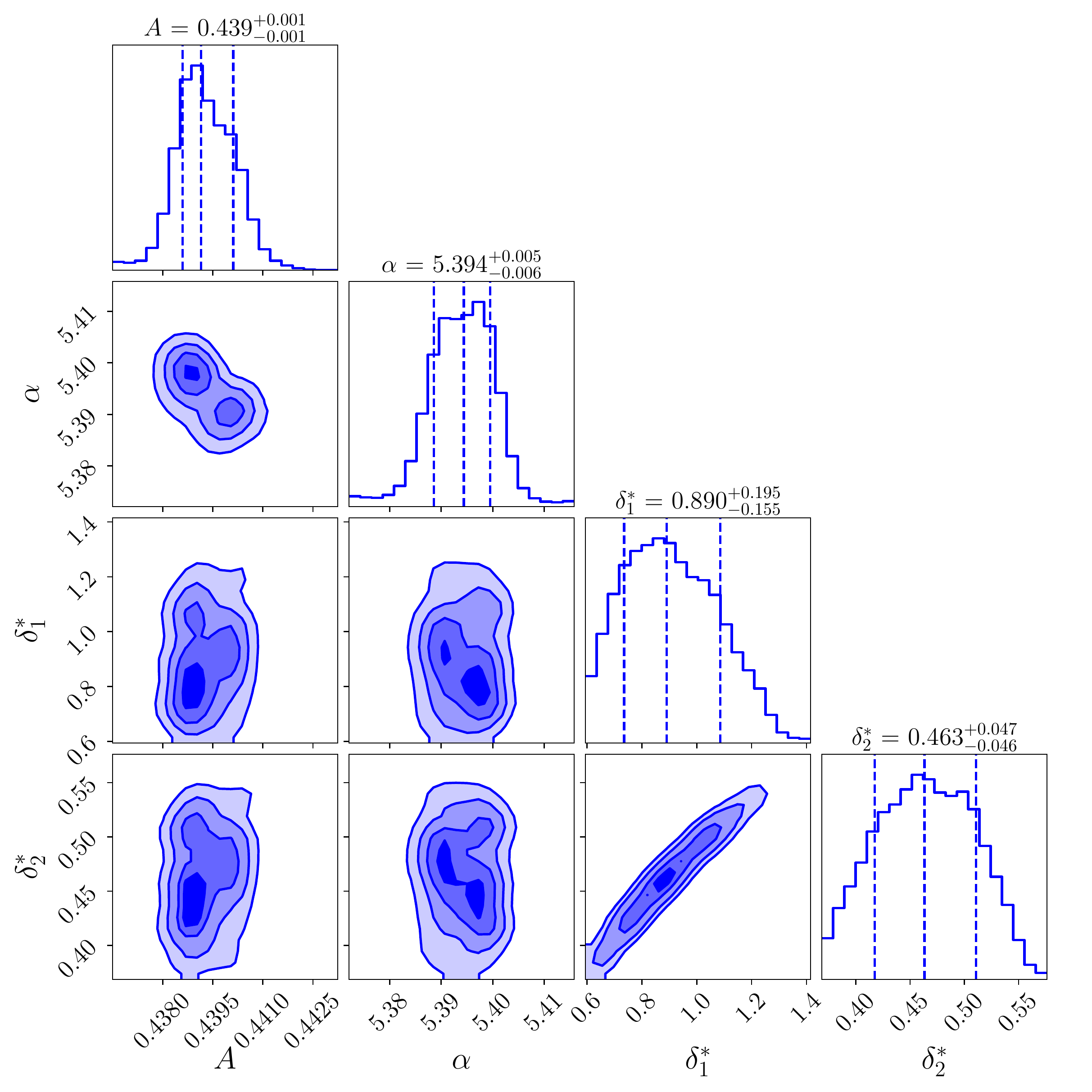}
    \caption{Posterior distributions of the $\{A,\, \alpha, \, \delta^*_1, \, \delta^*_2 \}$ free parameters, obtained from the partial fit of the \Lyaf{} PDF in knots.}
    \label{fig:posteriors}
\end{figure*}

\subsection{Fit of model parameters}

Given the model developed in \S\ref{sec:rsd} and \S\ref{sec:non-local}, we fit the coefficients of our model as follows. 

We first determine the RSD model parameters by finding the values which maximize the cross-correlation between the predicted and reference \Lyaf{} field (see the upper part of Table \ref{tab:table_pars}). In particular, we start with an initial guess based on the prior knowledge we have, i.e. we initialize the parameters $b_{v,i}=-0.8$, and $B_{i}=1$ $\forall i$, $i=\{{\rm knots,\,filaments,\, sheets, \, voids}\}$. As anticipated in \S\ref{sec:rsd}, we fix $\beta_i=0.5$ for all the cosmic web types, and set $B_{\rm sh}=B_{\rm vd}=0$. We then vary the parameters around the initial guess, one cosmic web at a time, apply the standard FGPA prediction, and monitor the variation of the cross correlation between our prediction and the reference \Lyaf{} field from the simulation. We notice that at this point we are still applying the standard FGPA, with the same parameters for all cells, because we have not yet fit the parameters for the non-local FGPA model. In order to be able to predict flux with the non-local FGPA at this point, one should perform an end-to-end automatic extraction of the parameters, going from the real-space dark matter field to the redshift-space non-local FGPA \Lyaf{} flux. However, this would imply a computationally expensive optimization procedure, involving interpolation of particles to the grid at each step. Therefore, we limit ourselves to a simpler manual determination which, while in principle suboptimal, turned out to be very instructive and allowed us to develop a good understanding of how different choices of the model parameters impact the cross correlation at different scales.  

\begin{figure*}
    \centering
    \includegraphics[width=\textwidth]{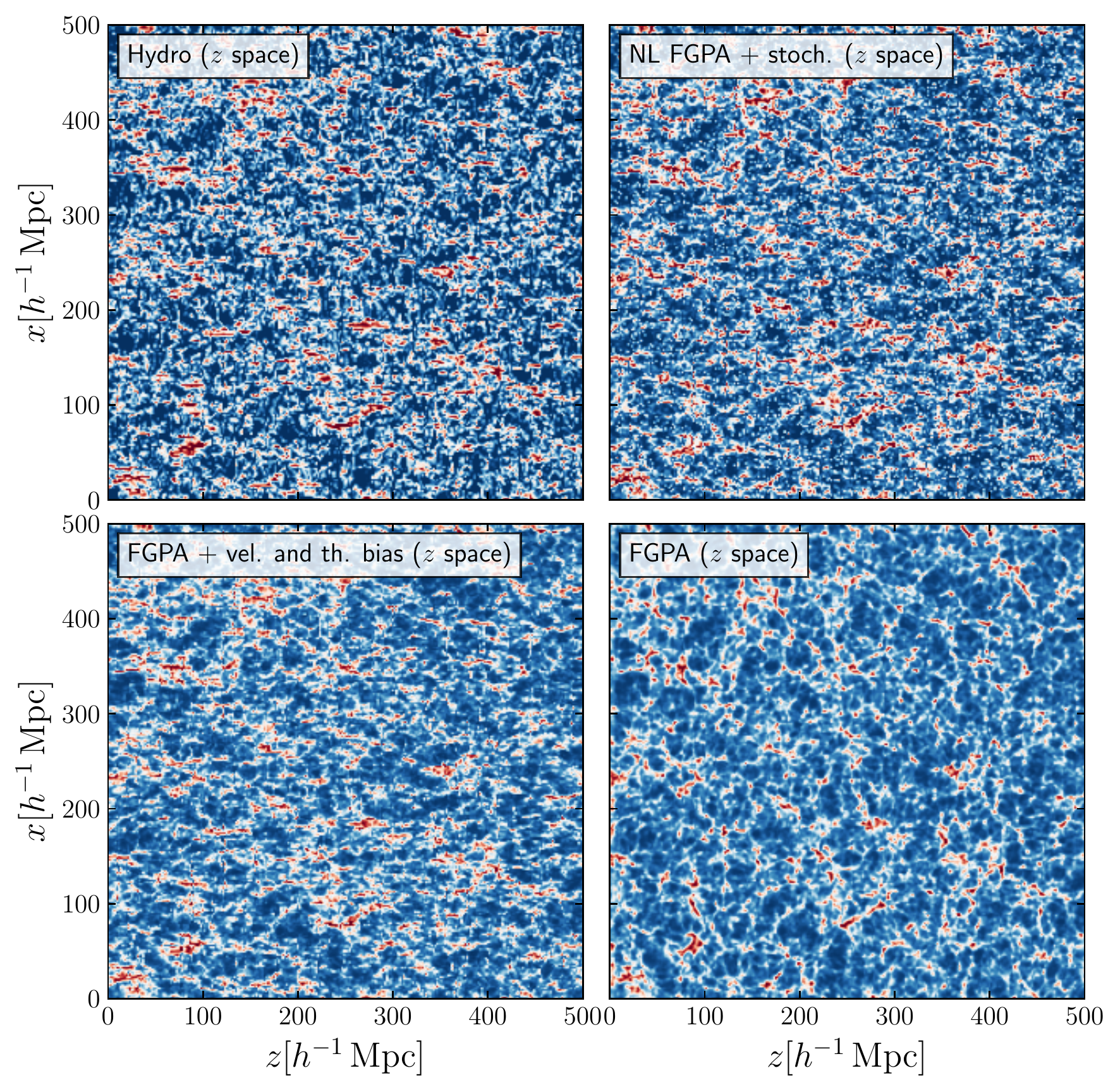}
    \caption{Slices through the simulation box, obtained by averaging two contiguous slices $1.95 \, h^{-1} \, {\rm Mpc}$ thick parallel to the $(x,z)$ plane, displaying redshift space distortions along the $z$ axis. The plot displays a \Lyaf{} transmitted flux $F/F_c$ slice through the reference cosmological hydrodynamic simulation (top left), and through the predicted \Lyaf{} boxes obtained through standard FGPA (bottom right), FGPA with velocity and threshold bias (bottom left), and our final non-local cosmic-web dependent FGPA and stochasticity (top right). The maps are color-coded from red to blue for values ranging in the interval $[0.1]$, where red indicates underdense and blue overdense regions. All the slices share the same color scale and the same extrema.}
    \label{fig:slices}
\end{figure*}

\begin{figure}
    \centering
    \includegraphics[width=\columnwidth]{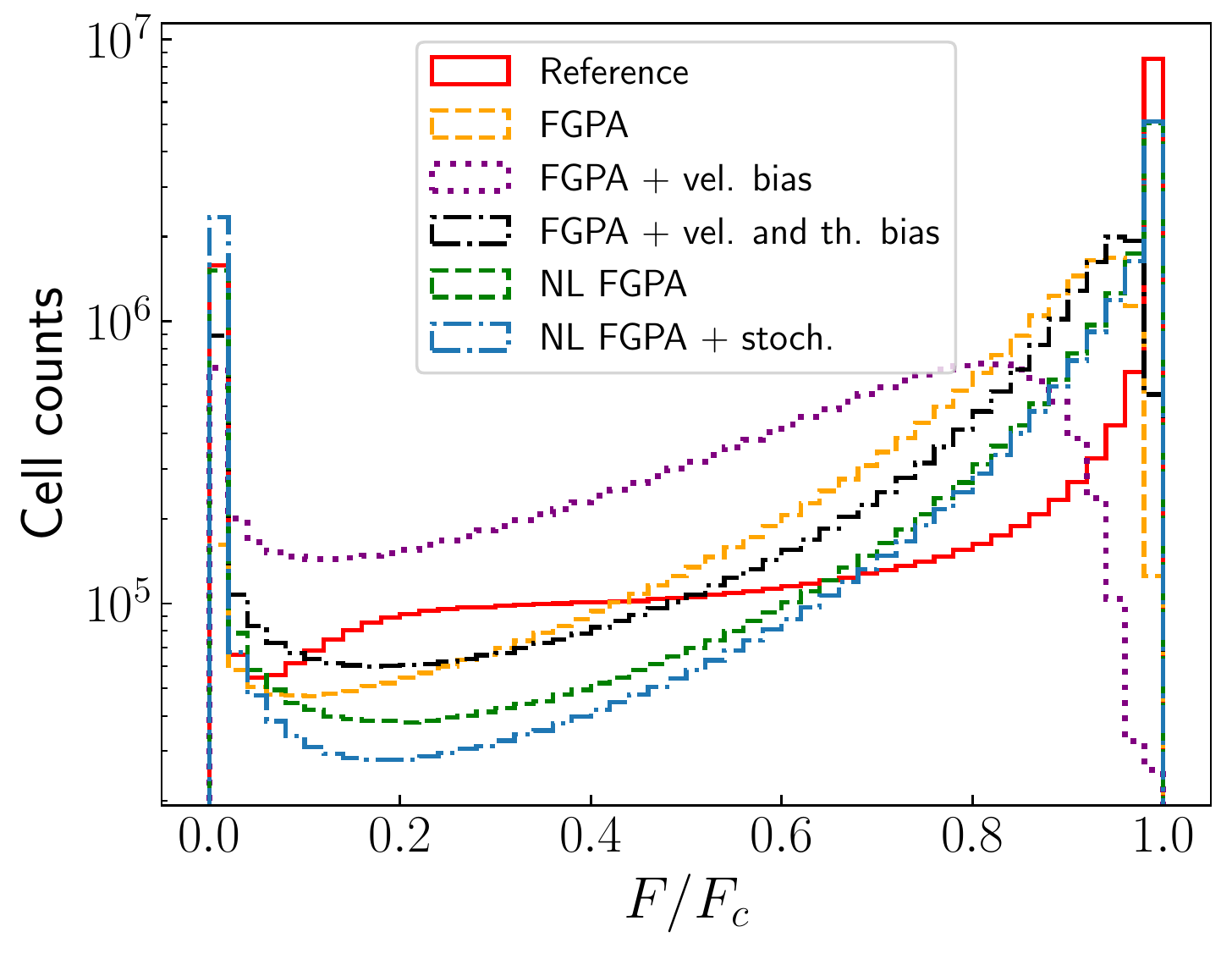}
    \caption{Comparison between probability distribution functions of transmitted \Lyaf{} flux $F/F_c$ as extracted from the reference simulation (red solid), and as predicted by the standard FGPA (yellow dashed), by the FGPA with velocity bias (purple dotted), by the FGPA with velocity and threshold bias  (black dashed-dotted), by the non-local FGPA(green dashed), and by the non-local FGPA and stochasticity (blue dashed-dotted).}
    \label{fig:pdf}
\end{figure}

Afterward, we determine the parameters for the modified non-local FGPA model. Here, we resort to automatic parameter estimation. 
In particular, we determine the optimal values for our parameters by sampling their posterior distribution through the affine-invariant \texttt{emcee} Markov Chain Monte Carlo (MCMC) sampler \citep{GoodmanWeare2010,ForemanMackey2013}.  

We proceed as follows. We separately fit the \Lyaf{} flux PDFs $\rho(F)$ of knots, filaments, sheets, and voids, i.e. performing a separate fit for each cosmic web type. In this way, each Markov Chain samples from the posterior distribution $p(\theta_i \,| \,{\rm data})\propto p({\rm ref \, | \, \theta_i}) \, p(\theta_i)$ of the parameters $\theta_i=\{A_i,\alpha_i,\delta^*_{1,i},\delta^*_{2,i}\}$, for $i=\{{\rm knots, \,filaments, \,sheets,\, voids}\}$, where "ref" denotes reference simulation \Lyaf{} PDF $\rho_{\rm ref}(F)$. 

We assume a Gaussian likelihood for $\rho_{\rm ref}(F)$:
\begin{equation}
    P(\rho(F)\, | \, \theta) =\prod_F \frac{1}{\sqrt{2\pi\sigma_F}} \, \exp \left[ -\frac{\left(\rho_{\rm ref}(F) - \rho_{\rm mock}(F)\right)^2}{2\sigma_F^2} \right],
\end{equation}
where $\rho_{\rm ref}(F)$ and $\rho_{\rm mock}(F)$ are the flux PDFs of the reference and predicted \Lyaf{} fields, and $\sigma_{F}$ corresponds to the number of cells containing a flux value inside the same bin of the PDF as $F$. Furthermore, we choose the following flat priors for the model parameters: $0<A_i<2, \, 1<\alpha_i<5, \, 0<\delta*_{1,i}<1.5, \, 0<\delta*_{2,i}<1.5, \, \forall i$. After running the chain with $32$ walkers for $2000$ iterations, we compute the autocorrelation length $\tau_{cl}$ using the in-built \texttt{emcee} function, and find that in all cases $\tau_{\rm cl}\sim 100$ iterations. Therefore, we conservatively discard the first $500$ iterations, i.e. $\sim 5$ times the chain autocorrelation length. As an example, we show the resulting posterior distribution of the $\{A,\, \alpha, \, \delta^*_1, \, \delta^*_2 \}$ free parameters obtained by fitting the \Lyaf{} PDF in knots. One clearly notices that there is a tight correlation between $\delta^*_1$ and $\delta^*_2$.

Eventually, we compute the median of the resulting posterior distribution, and the $16^{\rm th}$ and $84^{\rm th}$ percentiles intervals as associated uncertainties. 

Up to this point, the model does not include the stochastic noise term, and the predicted $P(k)$ suffers from a lack of power towards small scales with respect to the reference $P(k)$ (see \S\ref{sec:pk} for details). Therefore, the model parameters determined thus far correspond to the non-local FGPA without stochasticity. As described in \S\ref{sec:stochasticity}, we include a further additive noise term, exclusively in voids, randomly sampled from a negative binomial distribution. To determine the best parameters for the binomial distribution $\theta^\prime=\{f_\epsilon, n, p\}$, we first manually vary the parameters until we reach a combination of $\theta^\prime$ which enhances enough the small-scale power and provides a good fit of the power spectrum. (see \S\ref{sec:pk} for details). Eventually, we refit the parameters altogether, including the random component, and using as an initial guess the optimal parameter values previously found. Furthermore, we adopt the following upper and lower bounds as priors on $\theta^\prime$: $1<f_\epsilon<2$, $0<n<1$, $0<p<1$.

We report the final estimated parameter values and their uncertainty in Table \ref{tab:table_pars}. 
We find clear differences in the best-fitting values of the same parameter in different cosmic web environments. This supports our finding that including non-local terms is playing an informative role in the model about missing physical dependencies in the standard FGPA.

As a final note, we stress that only the fit of parameters for the negative binomial distribution governing the random sampling of the stochastic term was determined by considering deviations between the reference and predicted power spectrum. In fact, in the fit of the parameters of the non-local FGPA prescription, only the PDFs were taken into account. In this sense, the fact that this bias model reproduces the power spectrum and the bispectrum with such good accuracy constitutes a reassuring sanity check. 



\subsection{Visual inspection of slices through the simulation box} \label{sec:slices}

Figure \ref{fig:slices} shows a comparison of slices through the simulation box, obtained by averaging two contiguous slices $1.95 \, h^{-1} \, {\rm Mpc}$ thick parallel to the $(x,z)$ plane, displaying redshift space distortions along the $z$ axis. A visual inspection between the prediction from the standard FGPA (bottom right) and the reference cosmological simulation (top left) immediately reveals that the former fails to adequately render the redshift-space cosmic structures, displaying less pronounced redshift space distortions.  
When introducing a velocity bias correction to the standard FGPA (bottom left), redshift space distortions appear to model more appropriately the observed elongation of structures in the cosmic web, however, it turns out not to capture the saturation at $F\sim 1$ in the underdense regions, also reflected in the probability distribution function (see also Figure \ref{fig:pdf}). We again stress that the free parameters for the FGPA model have been chosen in this case in order to match the large-scale amplitude of the power spectrum. It would be also possible to find the parameters which best fit the PDF, but this yields a severe bias ($\gtrsim 30\%$) in the large-scale predicted power spectrum amplitude. Eventually, our non-local FGPA model\footnote{We hereafter omit to write that the non-local model includes also velocity and threshold bias.} (top right) visually resembles the reference simulation to a good degree of approximation. In particular, it succeeds both in reproducing the redshift-space structure of the cosmic web, and in properly predicting the right values of flux in all the different density regimes. Without the stochasticity, underdense regions in our predicted flux would appear emptier than they actually are. This aspect is reflected in a lack of power towards small scales, as will be commented more in detail in \S\ref{sec:pk}. The addition of the noise term $\epsilon$, only in voids, is meant to compensate for the lack of substructures. Therefore, we do not expect those small-scale structures to visually match the reference simulation in position. Rather, in a global sense, with the addition of the noise term voids appear to be more filled with substructures than in the case where we do not model stochasticity.   

\subsection{Probability distribution function}

Figure \ref{fig:pdf} shows a comparison between the \Lyaf{} PDF from the reference simulation (red solid), and the prediction from standard FGPA (dashed yellow), from FGPA with velocity bias (purple dotted), from FGPA with velocity and threshold bias (black dashed-dotted), and from our non-local FGPA without and with the addition of stochastic fluctuations in voids (green dashed and blue dashed-dotted, respectively).

The sharp bimodality of the \Lyaf{} PDF makes it particularly non-trivial to correctly predict both flux regimes, yielding the correct shape and average flux. While the standard FGPA model fails to reproduce the height of both peaks, the FGPA with velocity (and threshold) bias prescriptions achieve a better prediction towards $F\sim 0$, but a worse result at $F\sim 1$, as already commented in \S\ref{sec:slices}. In particular, the FGPA model with threshold bias achieves a better prediction of the PDF with respect to the FGPA model with only velocity bias, but none of the two is able to properly account for the low-density peak. The non-local FGPA, however, succeeds in reproducing both peaks, and qualitatively achieve the best result among the studied cases also in the intermediate flux regime $0.1\lesssim F \lesssim 0.9$. 

We stress that our final model, including stochasticity, with model parameters fitted to reproduce the PDFs of knots, filaments, sheets, and voids separately, yields a predicted average flux $\bar{F}_{\rm pred}\sim 0.75$, in great agreement with the value from the simulation $\bar{F}_{\rm ref}\sim 0.76$, even though the model has not been explicitly calibrated to reproduce such a quantity.

We point the reader's attention to the fact that the difference between the non-local FGPA (green dashed) and the FGPA model with velocity and threshold bias (black dashed-dotted) consists in making the model sensitive to non-local terms. It turns out to be clear that the non-local dependencies play an important role in improving the prediction of the \Lyaf{} PDF.

\begin{figure}
    \centering
    \includegraphics[width=\columnwidth]{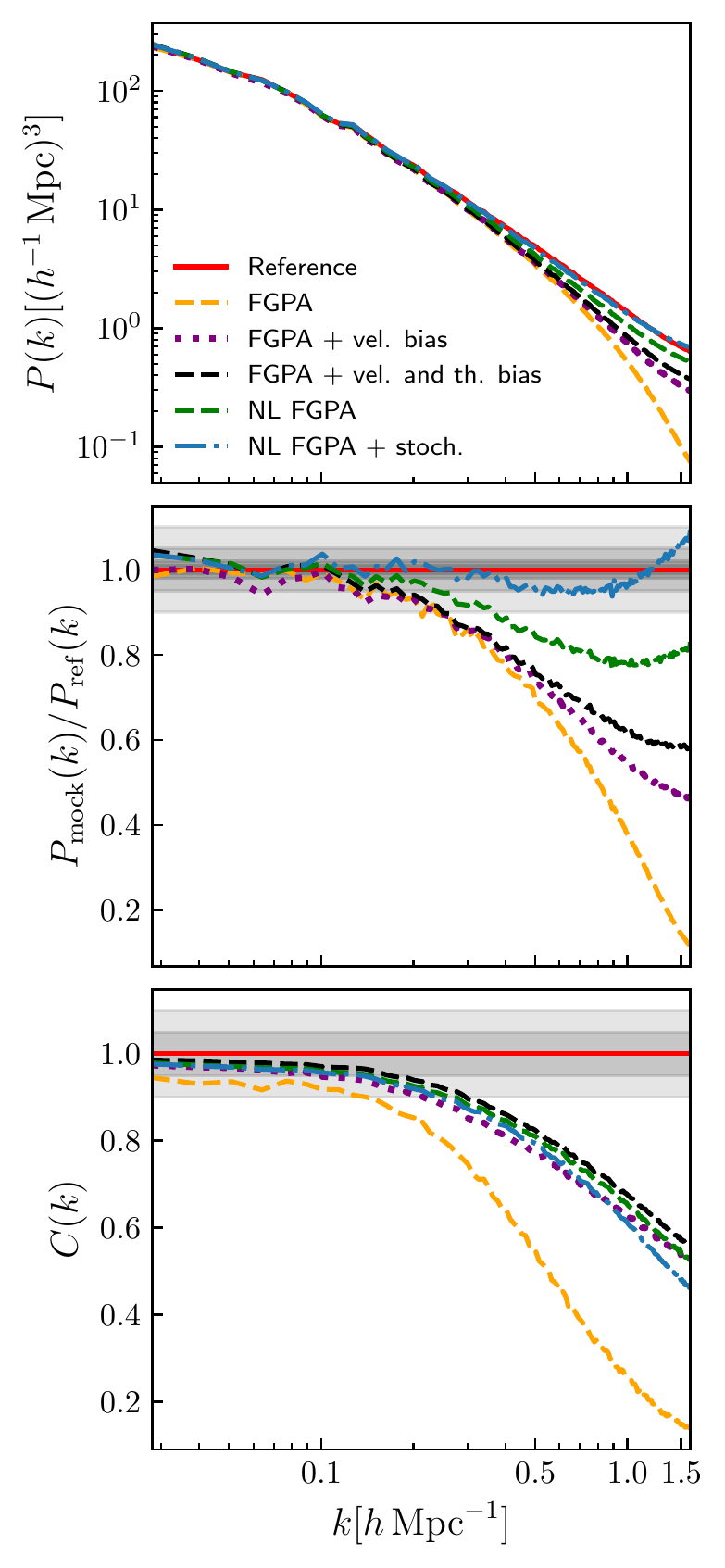}
    \caption{Top: comparison of 3D \Lyaf{} power spectrum $P(k)$. Mid: comparison of 3D \Lyaf{} power spectrum ratios $P_{\rm mock}(k)/P_{\rm ref}(k)$ between the predicted and reference power spectra. Bottom: comparison of cross-correlation coefficients $C(k)$ between the reference \Lyaf{} forest and the predictions from the different tested models. In each panel, the investigated reference simulation summary statistics are shown as a red solid line, while the prediction by the standard FGPA is displayed as a yellow dashed line, the FGPA with velocity bias as a purple dotted line, the FGPA with velocity and threshold bias as a black dashed-dotted line, the non-local FGPA as a green dashed line, and the non-local FGPA with stochasticity as a blue dashed-dotted line. In the mid panel, the gray shaded areas stand for $1\%,2\%,5\%$, and $10\%$ deviations, from darker to lighter. In the bottom panel, the gray shaded areas stand for $5\%$, and $10\%$ deviations, from darker to lighter.}
    \label{fig:pk}
\end{figure}

\subsection{Power spectrum and cross-correlation} \label{sec:pk}

In Figure \ref{fig:pk} we present a comparison between \Lyaf{} 3D power spectra $P(k)$ in the top panel, power spectrum ratios $P_{\rm mock}(k)/P_{\rm ref}(k)$ in the mid panel, and cross-correlation coefficients $C(k)$ in the bottom panel. We display the reference simulation summary statistics as a red solid line, while we plot the predicted \Lyaf{} flux summary statistics as a yellow dashed line in the case of standard FGPA, as a purple dotted line in the case of FGPA with velocity bias, as a black dashed-dotted line in the case of FGPA with velocity and threshold bias, as a green dashed line in the case of non-local FGPA, and as a blue dashed-dotted line the non-local FGPA with stochasticity. The top and bottom panels clearly highlight that the standard FGPA (yellow dashed) and its augmented version including velocity bias (purple dotted) and velocity and threshold bias (black dashed-dotted) rapidly lose power towards small scales, reaching ratios $P_{\rm mock}(k)/P_{\rm ref}(k)<20\%$, $P_{\rm mock}(k)/P_{\rm ref}(k)<50\%$, and $P_{\rm mock}(k)/P_{\rm ref}(k)<60\%$, respectively, at the Nyquist frequency $k_{\rm nyq}\sim 1.6 \, h \, {\rm Mpc}^{-1}$. As already anticipated, the $P(k)$ predicted by the non-local FGPA (green dashed) suffers from a $\sim 20\%$ lack of power towards small scales with respect to the reference $P(k)$. Eventually, the $P(k)$ yielded by the non-local FGPA model with stochasticity ensures a good fit of the target power spectrum, featuring maximum deviation $\sim 3\%$ and average deviations $\sim 0.1\%$ up to $k\sim 0.4 \, h \, {\rm Mpc}^{-1}$, maximum deviation $\sim 5\%$ and average deviations $\sim 1.8\%$ up to $k \sim 1.4 \, h \, {\rm Mpc}^{-1}$, and maximum deviation $\sim 8\%$ and average deviations $\sim 0.8\%$ up to the Nyquist frequency $k \sim 1.6 \, h \, {\rm Mpc}^{-1}$.

An inspection of the bottom panel of Figure \ref{fig:pk}, instead, reveals that the standard FGPA (yellow dashed) delivers a lower $C(k)$ with respect to all the other cases which incorporate the modeling of velocity bias. In fact, looking at Figure \ref{fig:slices}, it can be appreciated by eye  how the standard FGPA (top right) does not model properly the enhancement of cosmic structure elongation along the line of sight clearly seen in the reference simulation (top left). In this sense, an adequate treatment of velocity bias ensures a better cross-correlation at all scales. 
The non-local FGPA (green dashed) displays a larger $C(k)$ than the standard FGPA (yellow dashed), than the local FGPA with velocity bias (purple dotted) and than the local FGPA with velocity bias (purple dotted) at all scales, from a $1\%$ gain in $C(k)$ on large scales, up to a $\sim 3\%$ gain in $C(k)$ at $k\sim 1.0 \, h \, {\rm Mpc}^{-1}$. Eventually, the non-local FGPA with stochasticity (blue dashed-dotted) features a slightly lower $C(k)$ than the non-local FGPA (green dashed) up to $k\sim 0.5 \, h \, {\rm Mpc}^{-1}$, while it starts to depart from the latter at larger $k$, lacking a $\sim 6\%$ $C(k)$ at $k\sim 1.0 \, h \, {\rm Mpc}^{-1}$. This is expected by construction, due to the addition of a random component, which has the effect of lowering the cross correlation. However, the loss of $C(k)$ is not dramatic, and the non-local FPGA model with stochasticity (blue dashed-dotted) still improves the local FGPA with velocity bias $C(k)$ (purple dotted) up to $k\sim 0.8 \, h \, {\rm Mpc}^{-1}$. Remarkably, the FGPA model with velocity and threshold bias is found to provide the highest $C(k)$ across all scales, even though very similar to the prediction by the non-local FGPA model.

Here again, the difference between the non-local FGPA (green dashed) and the FGPA model with velocity and threshold bias (black dashed-dotted) consists in the inclusion of the non-local dependence on the cosmic web. The remarkable difference ($\gtrsim20\%$ gain in the accuracy of the predicted $P(k)$ at the Nyquist frequency) makes it clear the improvement accounted for by the non-local model.

\begin{figure*}
    \centering
    \includegraphics[width=\textwidth]{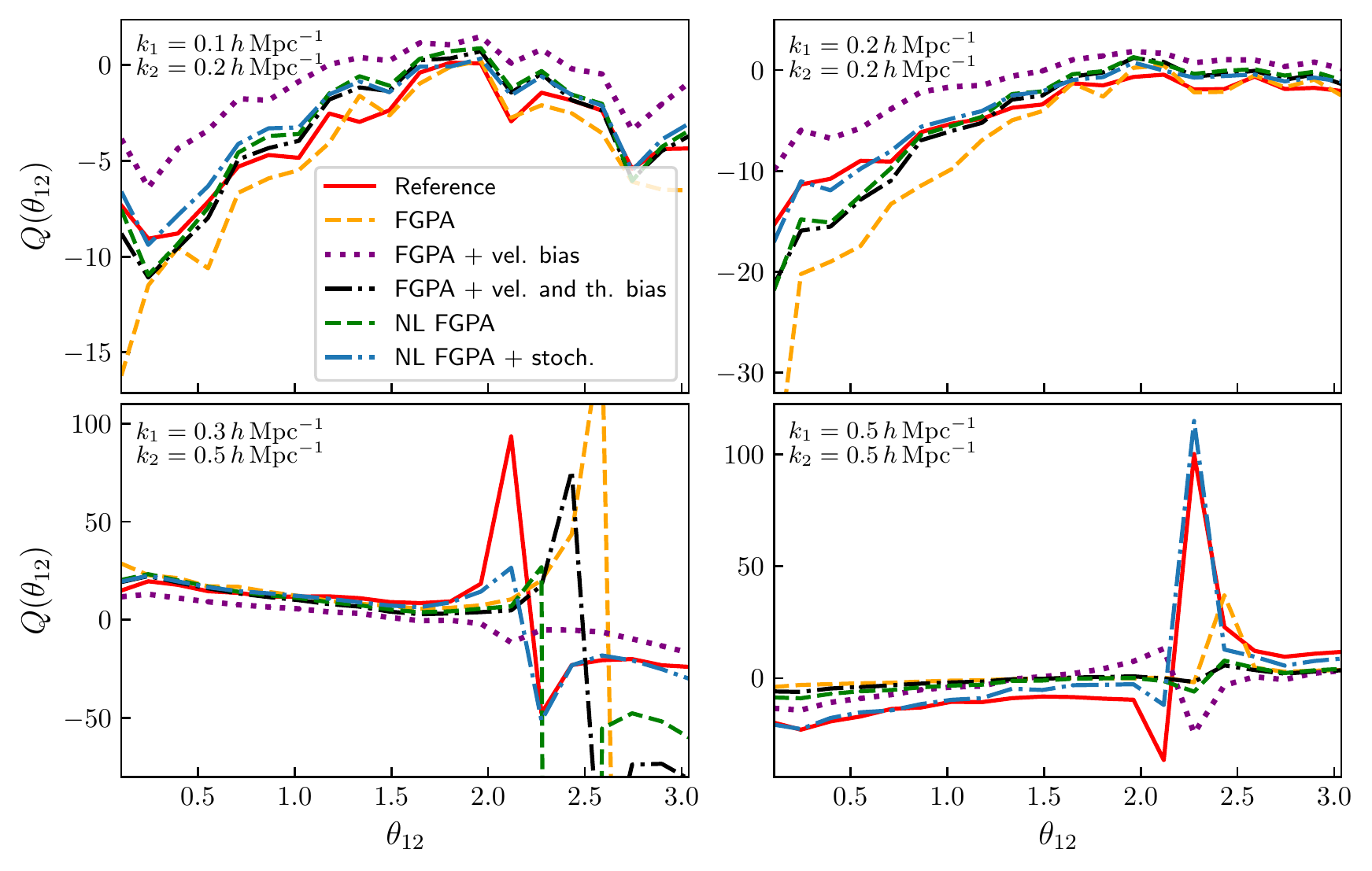}
    \caption{Reduced bispectrum $Q(\theta_{12})$ as a function of the subtended angle $\theta_{12}$, for four different triangular configurations: (i) $k_1=0.1,\, k=0.2 \, h \, {\rm Mpc}^{-1}$ (top left), (ii) $k_1=0.2,\, k=0.2 \, h \, {\rm Mpc}^{-1}$ (top right), (iii) $k_1=0.3,\, k=0.5 \, h \, {\rm Mpc}^{-1}$ (bottom left), (iv) $k_1=0.5,\, k=0.5 \, h \, {\rm Mpc}^{-1}$ (bottom right). The plot displays the bispectrum of the reference simulation (red solid), compared to the predictions from the standard FGPA (yellow dashed), the FGPA with velocity bias (purple dotted), the FGPA with velocity and threshold bias (black dashed-dotted),  the non-local FGPA (green dashed), and the non-local FGPA with stochasticity (blue dashed-dotted). The latter (i.e. our preferred model) reproduces the target bispectrum with reasonable accuracy for all the probed triangular configurations, while the others start to feature significant deviations at $k\gtrsim 0.2 \, h \, {\rm Mpc}^{-1}$.}
    \label{fig:bk}
\end{figure*}

\subsection{Bispectrum}

To address the capability of the model to reproduce higher-order statistics, we also assess the accuracy of the predicted bispectrum. We report in Figure \ref{fig:bk} the reduced bispectrum $Q(\theta_{12})$ as a function of the subtended angle $\theta_{12}$, for four different triangular configurations: (i) $k_1=0.1,\, k=0.2 \, h \, {\rm Mpc}^{-1}$ (top left), (ii) $k_1=0.2,\, k=0.2 \, h \, {\rm Mpc}^{-1}$ (top right), (iii) $k_1=0.3,\, k=0.5 \, h \, {\rm Mpc}^{-1}$ (bottom left), (iv) $k_1=0.5,\, k=0.5 \, h \, {\rm Mpc}^{-1}$ (bottom right). The plot displays the bispectrum of the reference simulation (red solid), compared to the predictions from the standard FGPA (yellow dashed), the FGPA with velocity bias (purple dotted), the FGPA with velocity and threshold bias (black dashed-dotted), the non-local FGPA (green dashed), and the non-local FGPA with stochasticity (blue dashed-dotted). The latter (i.e. our preferred model) reproduces the target bispectrum with reasonable accuracy for all the probed triangular configurations, while the others start to feature significant deviations at $k\gtrsim 0.2 \, h \, {\rm Mpc}^{-1}$. In particular, the non-local FGPA model with stochasticity reproduces not only the overall shape of the reference bispectrum, but also some peculiar features, such as the position and amplitude of the peak around $\theta \sim 2.3$ in the $k_1=k_2=0.5 \, h \, {\rm Mpc}^{-1}$ configuration, as well as the position (although not very much the amplitude) of the peak around $\theta \sim 2.1$ in the $k_1=0.3,\, k_2=0.5 \, h \, {\rm Mpc}^{-1}$.


\begin{figure}
    \centering
    \includegraphics[width=\columnwidth]{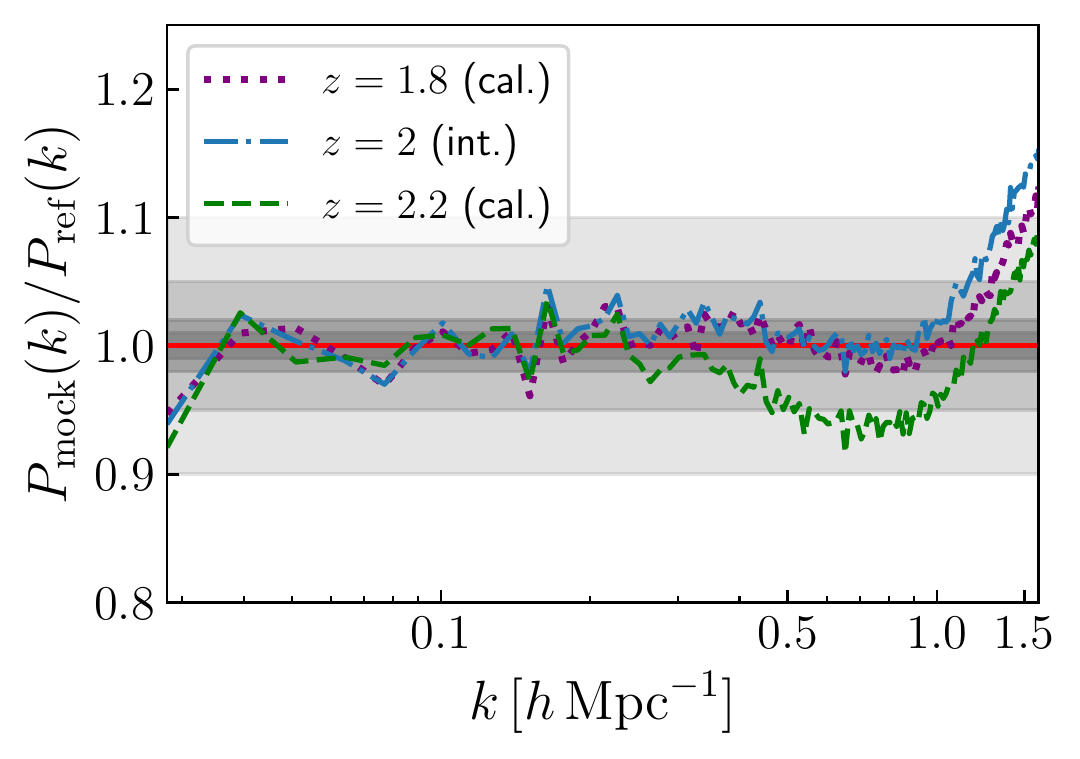}
    \caption{Comparison of 3D \Lyaf{} power spectrum ratios $P_{\rm mock}(k)/P_{\rm ref}(k)$ between the predicted and reference power spectra at different redshifts. Results at $z=1.8$ (purple dotted) and $z=2.2$ (green dashed) are obtained via independent calibrations. Instead, the $z=2$ result (blue dashed-dotted) corresponds to the \Lyaf{} power spectrum that we get by applying the non-local FGPA model with parameters linearly-interpolated at $z=2$ from calibrations at $z=1.8$ and $z=2.2$. The gray shaded areas stand for $1\%,2\%,5\%$, and $10\%$ deviations, from darker to lighter.}
    \label{fig:pk_interp}
\end{figure}


\section{Lightcone generation}\label{sec:lightcone}

In this section we illustrate how our method can be generalized to other redshifts and applied to generate \Lyaf{} lightcones.

We start by noticing that we have available from the simulation the following redshifts: 
$z=1.8,\,2.0,\, 2.2,\, 2.4,\, 2.6,\, 2.8,\, 3.0,\, \&\, 4.0$.

First, we show that the method generalizes also at other redshifts. To do so, we re-calibrate the parameters of our model also at $z=1.8$ and $2.2$. 
Figure~\ref{fig:pk_interp} shows the resulting \Lyaf{} power spectra ratios at $z=1.8$ (purple dotted) and at $z=2.2$ (green dashed). The achieved accuracy is very similar to the one shown in Fig. \ref{fig:pk} (mid panel) at $z=2$. Subsequently, we demonstrate that once the calibration is performed for some redshift snapshots, we can interpolate the parameters for the redshifts in between and ensure a continuous smooth redshift evolution. In particular, we linearly interpolate the parameters obtained at $z=1.8$ and $z=2.2$ and predict the \Lyaf{} field at $z=2$. We show the resulting $P(k)$ at $z=2$ in Fig. \ref{fig:pk_interp} (blue dashed-dotted), noticing that again the achieved accuracy is very similar to the one we get via direct calibration at $z=2$.

\begin{figure*}
    \centering
    \includegraphics[width=\textwidth]{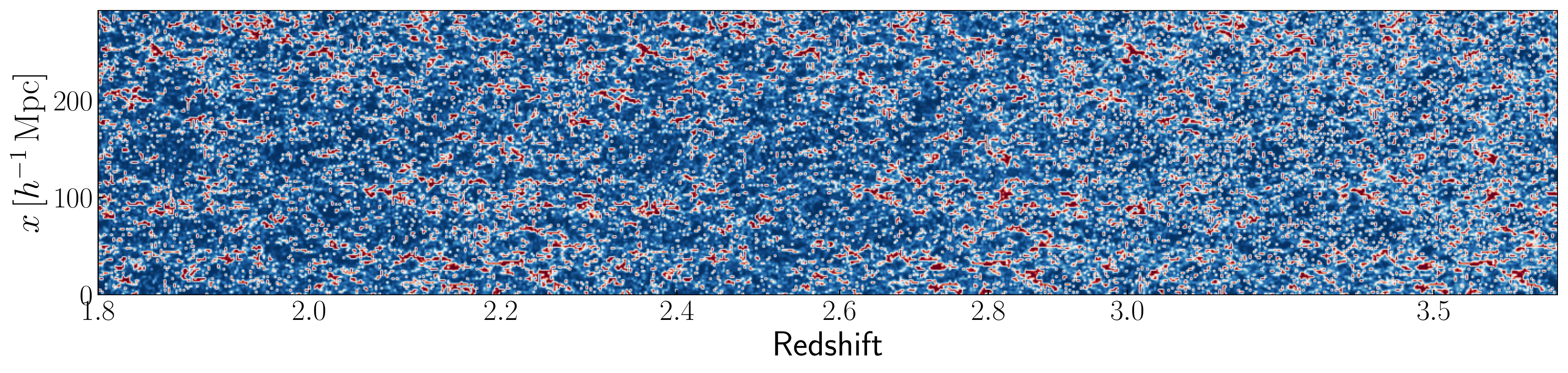}
    \caption{Slice through the lightcone box with redshift evolution from $z=1.8$ to $z=3.8$. The lightcone is obtained by replicating three times the same simulation box along the redshift direction and connecting the different redshift space dark matter field snapshots. Afterwards, the \Lyaf{} was painted on the dark matter lightcone by interpolating the parameters of the model calibrated at different redshift snapshots, as illustrated in \S\ref{sec:lightcone}. Once can visually appreciate the redshift evolution, with voids becoming bigger and emptier towards the low-redshift end.}
    \label{fig:lightcone}
\end{figure*}

Afterwards, we re-calibrate the model parameters also on the remaining redshift snapshots and generate a lightcone box as follows. We consider the redshift range $\Delta z=[1.8,\, 3.8]$. The comoving radial distance subtended by this redshift interval is $l\sim 1500 \, h^{-1}\, {\rm Mpc}$. Therefore, after having mapped the dark matter fields at all redshift from real to redshift space, we replicate three times the same simulation box along the redshift direction. Then, we connect the redshift space dark matter field snapshots transitioning from one to the following at the average redshift between the two. Even though this implies a sharp transition in the dark matter field, we paint the \Lyaf{} on the lightcone by interpolating the parameters of the model with a cubic spline interpolation scheme at each redshift slice. This ensures a smooth redshift evolution. We show a slice through the final lightcone box in Fig. \ref{fig:lightcone}. One can clearly visualize how the \Lyaf{} changes from $z=1.8$ to $z=3.8$, with voids becoming bigger and emptier towards the low-redshift end.

In a forthcoming publication, we will present full-sky \Lyaf{} mocks by generating dark matter fields with lightcone geometry and smooth evolution in redshift already in the dark matter field (Sinigaglia, Kitaura et al. in prep.).

\section{Potential future improvements} \label{sec:outlooks}

While the presented model, as anticipated, turns out to be sufficient to fulfill the purpose it was designed for, there are further potential improvements which we leave to be explored in future works.

In this work, we have included non-local dependencies in the bias formulation only through the T-web. Such dependencies are known to have a long-range effect \citep[e.g.][]{McDonaldRoy2009}, and we are hence neglecting short-range terms, whose effect kicks in when it comes to modeling the non-linear clustering towards small scales. Short-range non-local terms can be constructed as scalars from the curvature tensor $\delta_{ij}=\partial_i\partial_j\delta$, such as its Laplacian $\nabla^2\delta$ in Eulerian \citep[e.g.][]{Peacock1985,McDonaldRoy2009,WernerPorciani2020,Kitaura2022} or in Lagrangian space \citep[e.g.][]{Zennaro2022}, which characterizes the shape of the local maxima of the density field \citep{Peacock1985}. More in general, one can build short-range non-local terms as scalars computed from arbitrarily higher-order derivatives of the density field, such as $\partial^l_i\partial^l_j\delta$ \citep[e.g.][]{Kitaura2022}. Alternatively, in analogy with the I-web description based on the invariants of the tidal field tensor (see also \S\ref{sec:cwc}), one can work in a framework relying on the invariants of the curvature tensor. This latter description has been proven to encode crucial information to model the bias of baryon density fields \citep[e.g.][]{Sinigaglia2021,Sinigaglia2022}. Since we do not model such family of dependencies here, we stress that we may be missing some relevant piece of physical information. However, we also stress that computing such terms requires an accurate enough modeling of small scales, which is not guaranteed here. Therefore, by introducing short-range non-local terms we may face the risk of introducing a noisy component, which goes in detriment of the final accuracy. Moreover, in the computationally cheapest way of modeling such dependencies, we would be extracting the analogous of the T-web based on $\delta_{ij}$, which would imply a non-negligible number of additional free parameters in our model.     

In connection to the previous point, we notice that \cite{Kitaura2022} showed that the I-web model encodes a larger amount of information on the clustering of biased tracers with respect to the T-web, at the expense of a larger number of parameters (the number of bins used to described the invariants) and therefore at a larger risk to incur into overfitting. Moreover, while describing the non-local quantities by explicitly using the invariants of the tidal tensor is possible \citep[e.g.][]{Pellejero2020}, it implies devising a suitable functional form for each additional term used in the bias, which is in principle non-trivial.

Another point which can be improved consists in the way we model the low-density regions, identified as cosmic voids according to the T-web classification. In fact, in order to compensate the lack of substructures, we randomly sample a noise component from a negative binomial distribution. At this point we notice that this could be avoided, or its impact alleviated, by improving the modeling of the deterministic component. One easy modification could consist in binning the density distribution in voids into an arbitrary number of intervals, and identify distinct parameters for such distinct density bins. However, this would again introduce a larger number of free parameters in our model.

Eventually, if a noise term is to be included as done in this work, we notice that we have randomly sampled the stochastic component from a negative binomial distribution for the reasons previously stated, but this is not the only possible solution. In fact, one may find that using a different distribution may be more convenient.

We leave all these potential improvements to be investigated in future publications, or in applications adopting this model.


\section{Conclusions}\label{sec:conclusions}

This work presents a significantly augmented version of the widely-used FGPA to predict the \Lyaf{} in redshift space, as observed from current and forthcoming spectroscopic galaxy surveys. This new model relies on explicit modeling of velocity bias and redshift space distortions and of a modification to the FGPA introducing a cut-off and a boosting scale, with free parameters (determined based on both heuristic and physical arguments, as well as on an efficient MCMC scheme) made dependent on the cosmic web environment as described by the T-web classification \citep[knots, filaments, sheets, and voids, ][]{Hahn2007}. In this sense, we label our model \textit{non-local} FGPA. In fact, the \Lyaf{} flux in each cell is here made dependent on the content of the surrounding cells and the geometry of the gravitational field, and the model effectively incorporates non-local bias information up to third-order in Eulerian perturbative bias expansion \citep[see][]{Kitaura2022}. In addition, since we find that such a model fails to reproduce the small-scale clustering in the underdense regions, we add a further stochastic term only in voids, randomly sampled from a negative binomial distribution. This random component has the effect of enhancing the small-scale power, and hence, of improving the fit of the \Lyaf{} forest 3D power spectrum towards small scales.

We predict \Lyaf{} fluxes with standard FGPA, our preferred non-local FGPA model with stochasticity, and three other intermediate models in between, on a mesh with $V=(500 \, h^{-1}\,{\rm Mpc})^3$ volume and $N_{\rm c}=256^3$ cells, with physical cell resolution $l\sim 1.95 \, h^{-1}\,{\rm Mpc}$, trying to reproduce a full cosmological hydrodynamic N-body/SPH simulation (spanning the same volume) and run with $N=1024^3$ particles. We assess the accuracy of the prediction of such models by comparing the results regarding the mean transmitted flux $\bar{F}$, the PDF, the power spectrum, and the bispectrum, with the analogous summary statistics computed from the reference simulation.  

The augmented non-local FGPA with stochasticity improve upon the standard FGPA model in all the investigated summary statistics. In fact, the non-local FGPA accurately reproduces the mean transmitted flux $\bar{F}$, the flux PDF, the power spectrum, and the bispectrum. In particular, our model yields a  mean transmitted flux $\bar{F}_{\rm pred}\sim 0.75$, in excellent agreement with the value $\bar{F}_{\rm ref}\sim 0.76$, and a power spectrum featuring maximum deviation $\sim 3\%$ and average deviations $\sim 0.1\%$ up to $k\sim 0.4 \, h \, {\rm Mpc}^{-1}$, maximum deviation $\sim 5\%$ and average deviations $\sim 1.8\%$ up to $k \sim 1.4 \, h \, {\rm Mpc}^{-1}$, and maximum deviation $\sim 8\%$ and average deviations $\sim 0.8\%$ up to the Nyquist frequency $k \sim 1.6 \, h \, {\rm Mpc}^{-1}$. The predicted PDF and the bispectrum clearly outperform the results from any other model tested in this paper as well.  

Compared to other schemes based on machine learning \citep[e.g.,][]{Harrington2021,Horowitz2021,Sinigaglia2022}, and other methods based on iterative calibrations \citep[e.g.,][]{Sorini2016,Peirani2022}, our model offers the appeal of being a purely analytical method, and therefore it is fast to compute, less prone to the overfitting problem, and it can be straightforwardly generalized to larger/smaller volumes, different mesh resolutions, and at different redshift snapshots. This method can even ensure a continuous smooth treatment of redshift evolution via interpolation of the fitting coefficients in between the redshift snapshots used for calibration, avoiding discontinuity problems at the edge of distinct contiguous redshift shells.

We point out that the resolution ($1.95 \, h^{-1} \, {\rm Mpc}$ physical cell size) considered in this work is not sufficient to provide realistic \Lyaf{} high-resolution spectra with correct modeling of the 1D power spectrum, needed to perform studies of small-scale ($k\sim5-10\,h\,\rm{Mpc}^{-1}$) Physics, such as  bounds on the mass of warm dark matter particles \citep{Viel2005,Viel2013a} and of neutrinos \citep{PalanqueDelabrouille2015} and constraints on the thermal state of the intergalactic medium \citep[][and references therein]{Garzilli2017}, among others. While the application of the non-local FGPA to high-resolution density fields is possible, one should bare in mind that resolving sub-Mpc scales on a 3D mesh makes it computationally very expensive to realize large-volume cosmological boxes. Therefore, in future works, we will explore novel techniques based on Bayesian inference aimed at upsampling the resolution of 1D \Lyaf{} skewers extracted from low-resolution \Lyaf{} flux boxes generated following the procedure presented in this work.

We plan to adopt this novel scheme to generate \Lyaf{} mock lightcones for the DESI survey, in combination with accurate approximated gravity solvers correcting for shell-crossing \cite[e.g.][]{Kitaura2013,Tosone2021,Kitaura2023}. In particular, we aim at using a recent fast structure formation model based on iterative Eulerian Lagrangian Perturbation Theory \citep[eALPT,][]{Kitaura2023}, which has been shown to reproduce the clustering of N-body simulations with percent accuracy deep into the non-linear regime. Furthermore, the application of the non-local FGPA model to approximated dark matter fields generated through the ALPT or eALPT approximations can potentially represent a new avenue in Bayesian density field reconstruction studies using the \Lyaf{}. We will investigate this point in future publications.

In conclusion, we argue that this novel non-local FGPA scheme represents a significant step forward with respect to previous analytical efforts, and it can become instrumental in the generation of fast accurate mocks for covariance matrices estimation in the context of current and forthcoming \Lyaf{} surveys. 

The code implementing the model presented in this work is made publicly available.\footnote{\url{https://github.com/francescosinigaglia/nl-fgpa}}


\begin{acknowledgements}
      The authors wish to warmly thank the anonymous referee for the insightful comments they provided, which helped to improve the quality of the manuscript.
      We are grateful to Cheng Zhao for making his bispectrum computation code available. F.S. acknowledges the support of the doctoral grant funded by the University of Padova and by the Italian Ministry of Education, University and Research (MIUR) and the financial support of the \textit{Fondazione Ing. Aldo Gini} fellowship. F.S. is indebted to the \textit{Instituto de Astrofísica de Canarias} (IAC) for hospitality and the availability of computing resources during the realization of this project. F.S.K. and A.B.A acknowledge the IAC facilities and  the Spanish Ministry of Science and Innovation (MICIN) under the Big Data of the Cosmic Web project PID2020-120612GB-I00.   
      K.N. is grateful to Volker Springel for providing the original version of {\sc GADGET-3}, on which the {\sc GADGET3-Osaka} code is based. Our numerical simulations and analyses were carried out on the XC50 systems at the Center for Computational Astrophysics (CfCA) of the National Astronomical Observatory of Japan (NAOJ), the Yukawa-21 at YITP in Kyoto University, {\sc SQUID} at the Cybermedia Center, Osaka University as part of the HPCI system Research Project (hp210090).  This work is supported in part by the JSPS KAKENHI Grant Number JP17H01111, 19H05810, 20H00180 (K.N.), 21J20930, 22KJ2072 (Y.O.). K.N. acknowledges the travel support from the Kavli IPMU, World Premier Research Center Initiative (WPI), where part of this work was conducted. 
\end{acknowledgements}

%
%


\bibliographystyle{aa}
\bibliography{lit}




   
  



\end{document}